\documentclass[12pt]{iopart}

\usepackage{amssymb}

\expandafter\let\csname equation*\endcsname\relax
\expandafter\let\csname endequation*\endcsname\relax
\usepackage{amsmath}
\usepackage[pdfborder=0,colorlinks=true,linkcolor=black,citecolor=black]{hyperref}
\usepackage[all]{hypcap}

\newcommand{\be}{\begin{equation}}
\newcommand{\ee}{\end{equation}}

\def\link{\prec\!\!*\,}
\def\CS{\mathcal{C}}
\def\mink{\mathbb{M}}

\def\LP{L\{P\}}
\def\Bbar{\bar{B}}

\begin{document}

\title{Correction terms for propagators and d'Alembertians due to spacetime discreteness}
\author{Steven Johnston}
\ead{steven.p.johnston@gmail.com}

\begin{abstract}
The causal set approach to quantum gravity models spacetime as a discrete structure --- a causal set. Recent research has led to causal set models for the retarded propagator for the Klein-Gordon equation and the d'Alembertian operator. These models can be compared to their continuum counterparts via a sprinkling process. It has been shown that the models agree exactly with the continuum quantities in the limit of an infinite sprinkling density --- the continuum limit. This paper obtains the correction terms for these models for sprinkled causal sets with a finite sprinkling density. These correction terms are an important step towards testable differences between the continuum and discrete models that could provide evidence of spacetime discreteness.
\end{abstract}


\submitto{\CQG}

\maketitle

\section{Introduction}

Causal set theory is an approach to quantum gravity in which the continuum manifold description of spacetime is replaced by a discrete description based on a causal set. A causal set is a locally finite partially ordered set in which the set elements represent spacetime events and the  partial order represents the causality relations between them. See \cite{Bombelli:1987, Sorkin:2003,Henson:2006, Surya:2011} for full details.

In recent years, causal set models have been developed for the retarded and Feynman propagators of the Klein-Gordon equation \cite{Johnston:2008, Johnston:2009} and the d'Alembertian operator \cite{Sorkin:2008, Benincasa:2010, Dowker:2013vba,Glaser:2014, Aslanbeigi:2014}. These models are defined implicit to a causal set but their behaviour can be compared to their continuum counterparts when the causal sets are generated by a sprinkling process from a Lorentzian manifold. Sprinkling is a random process that depends on a single parameter, the sprinkling density, that measures the expected number of causal set elements that will be sprinkled into a spacetime volume.

The parameters in the models are chosen so that their expectation value, averaged over sprinkled causal sets, agrees with the corresponding continuum quantity for causal sets generated by sprinkling into Minkowski spacetime. In most instances, the agreement is exact only in the limit of infinite sprinkling density (i.e. the continuum limit). The results of this paper provide the correction terms for large, finite sprinkling densities.

These correction terms are an important step towards phenomenological results from causal set theory in which the effects of a fundamental spacetime discreteness could be detected. They also provide an analytical approach to analyse the behaviour of the models for very large sprinkling densities, something that is difficult to do through direct computer simulations.

An alternative view of causal sets is as Lorentz-invariant discretizations of Lorentzian manifolds. In this view, the correction terms derived here may offer a natural cut-off to tame the divergences in continuum quantum field theories.

\subsection{Preliminaries}
\label{sec:Preliminaries}

A \emph{causal set} (or \emph{causet}) is a locally finite partial order, i.e.
a pair $(\CS,\preceq)$ where $\CS$ is a set and $\preceq$ a relation on $\CS$ which is (i) reflexive ($x\preceq x$); (ii) antisymmetric ($x \preceq y \preceq x \implies  x = y$); (iii) transitive ($x \preceq y \preceq z \implies x \preceq z$); and (iv) locally finite ($\left| [x,y] \right| < \infty$) for all $x, y\in \CS$. Here $[u,v]:=\{w \in \CS : u \preceq w \preceq v\}$ is a causal interval and $\left| A \right|$ denotes the cardinality of a set $A$. We write $x \prec y$ if $x \preceq y$ and $x \neq y$. A \emph{link} between $u, v \in \CS$ (written $u \link v$) is a relation $u \prec v$ such that there exists no $w \in \CS$ with $u \prec w \prec v$. A finite \emph{chain} of length $n$ is a sequence of distinct elements $u_0 \prec u_1 \prec u_2 \prec \ldots \prec u_n$. A finite \emph{path} of length $n$ is a sequence of distinct elements $u_0 \link u_1 \link u_2 \link \ldots \link u_n$.

The set $\CS$ represents the set of spacetime events and the partial order $\preceq$ represents the causal order between pairs of events. If $x \preceq y$ we say ``$x$ is to the causal past of $y$''. The causal relation of a Lorentzian manifold (without closed causal curves) satisfies conditions (i)-(iii). It is condition (iv) that enforces spacetime discreteness---each causal interval contains only a finite number of events.

\emph{Sprinkling} is a way to generate a causal set from a $d$-dimensional Lorentzian manifold $(M,g)$. Points are placed at random within $M$ using a Poisson process (with sprinkling density $\rho$) so the expected number of points in a region of $d$-volume $V$ is $\rho V$. This generates a causal set whose elements are the sprinkled points and whose partial order relation is ``read off'' from the manifold's causal relation restricted to the sprinkled points.

Here we shall only consider causal sets generated by sprinkling into $d$-dimensional Minkowski spacetime, $\mink^d$. To fix notation, we let $x$ be a point in $\mink^d$ with coordinates $(x_0,x_1,\ldots,x_{d-1})$ and use the quadratic form $\omega(x) = x^2 = x_0^2 - x_1^2 - \ldots - x_{d-1}^2$. With this metric signature the d'Alembertian operator is 
\be \Box = \frac{\partial^2}{\partial x_0^2} - \frac{\partial^2}{\partial x_1^2} - \frac{\partial^2}{\partial x_2^2} \ldots - \frac{\partial^2}{\partial x_{d-1}^2} \ee
The causal partial order for $x, y \in \mink^d$ is defined as: $x \preceq y \iff x_0 < y_0 \textrm{ and } (x-y)^2 \geq 0$.
The solid forward light cone at a point $x \in \mink^d$ is the set $\Gamma^+_x = \{y \in \mink^d: x \preceq y\}$

Quantities calculated on a sprinkled causal set can be averaged over multiple sprinklings (with the same sprinkling density $\rho$) to give a function in $\mink^d$. The behaviour of this expectation value allows the discrete and continuum theories to be compared. We shall focus on the expectation values for the retarded propagator and the d'Alembertian operators for sprinklings into $\mink^d$. The behaviour of these expectation values for large, finite values of $\rho$ will be examined.

\section{Chains and Paths}

To begin it will be useful to look at the expected number of chains and paths in a causal set generated by sprinkling into $\mink^d$ with sprinkling density $\rho$. To assist us, we define the following important functions:
\be \label{eq:Cdef} C(y-x) := \left\{ \begin{array}{ll} 1 & \textrm{ if } x \preceq y \\ 0 & \textrm{otherwise,}\end{array} \right.\ee
\be \label{eq:Vdef} V(x) = C(x) c_d  (x_0^2 - x_1^2 - \ldots - x_{d-1}^2)^{d/2} \quad \textrm{with}\quad  c_d := \frac{\pi^{\frac{d-1}{2}}}{2^{d - 1} d \,\Gamma\left(\frac{d + 1}{2}\right)} \ee
\be \label{eq:Pdef} P(x) := C(x) e^{-\rho V(x)} \ee
The function $C(x)$ is the ``causal step function'' --- the indicator function for the closed forward light cone $\Gamma^+_0$. The function $V(y-x)$ is the $d$-dimensional volume of the causal interval $[x,y] = \{z \in \mink^d : x \preceq z \preceq y\}$ and the function $P(y-x)$ denotes the probability that the sprinkled causal set has a link between $x$ and $y$.

For two functions $f, g$ on $\mink^d$ their convolution is defined as:
\be (f*g)(x):= \int_{\mink^d} f(x-y) g(y) \mathrm{d}^dy. \ee

The expected number of chains of length one from $x$ to $y$ is $C_1(y-x) := C(y-x)$. The expected
number of chains of length $n > 1$ from $x$ to $y$ is given by
\be \label{eq:ChainDef} C_n(y-x) := \rho (C*C_{n-1})(y-x) = \rho^{n-1}(C*C*\ldots *C)(y-x),\ee
where there are $n$ copies of $C$ and $n-1$ convolutions in the final expression. From the definition we have $\rho C_n * C_m = C_{n+m}$.

The expected number of paths of length one from $x$ to $y$ is given by $P_1(y-x) := P(y-x)$. The expected number of paths of length $n > 1$ from $x$ to $y$ is given:
\be \label{eq:Pn} P_n(y-x) := \rho(P * P_{n-1})(y-x) = \rho^{n-1}(P * P * \ldots * P)(y-x),\ee 
where there are $n$ copies of $P$ and $n-1$ convolutions.

We note that these functions can be used to define regular distributions through convolution. For example, $C_n$ defines a distribution for suitable test functions $f$ with action $\langle C_n, f \rangle = C_n * f$.

We recall the known results for $C_n(x-y)$, calculated explicitly in \cite[Theorem III.2, p50]{Meyer:1988}. For $n \geq 1$ we have:
\be \label{eq:ChainForm} C_n(y-x) = C(y-x) (\rho V(x-y))^{n-1} D_n, \ee
where $D_1:=1$ and
\be D_n:=\frac{1}{n-1} \left( \frac{\Gamma(d+1)}{2}\right)^{(n-2)} \frac{\Gamma(d/2) \Gamma(d)}
{\Gamma((n-1)d/2) \Gamma(n d/2)}, \ee
for $n > 1$ is a real dimensionless constant.

To the author's knowledge, there is no known closed-form expression for $P_n(y-x)$. Power series based on recurrence relations have been obtained in \cite[Sec A.2]{Johnston:2010} and an approximation for the total number of paths between $x$ and $y$ in $d=2$ was derived in \cite[eq 2.5.15]{Bombelli:1987:PhD}.

Useful insight into the function $P(x)$ can be obtained by expanding it in terms of the $C_n(x)$ functions to give:
\be \label{eq:PintermsofC} P(x) = C(x) e^{-\rho V(x)} = C(x) \sum_{n=0}^\infty \frac{(-\rho V(x))^n}{n!} = \sum_{n=0}^\infty \frac{(-1)^n C_{n+1}(x)}{n! D_{n+1}},\ee
where we have used \eqref{eq:ChainForm}.

We can now apply the d'Alembertian operator using the observation that for even dimensions $d$ we have \cite[eq 3.4 with $k=d/2$]{Kolk:Riesz}:
\be \Box^{d/2} C(x) = A \delta(x)\quad \textrm{with}\quad A = 2^{d-1} \pi^{(d-2)/2} \Gamma(d/2)\ee
where $\delta(x)$ is the $d$-dimensional delta-function and $A$ is a constant.
From this and \eqref{eq:ChainDef} we have
\be \Box^{d/2} C_n = \rho \left(\Box^{d/2} C\right) * C_{n-1} = A \rho \delta * C_{n-1} =  A \rho C_{n-1}. \ee
Combining with \eqref{eq:PintermsofC} it immediately follows that 
\be \label{eq:BoxP} \Box^{d/2} P = A \delta + A \rho \sum_{n=1}^\infty \frac{(-1)^n C_{n}}{n! D_{n+1}}  =  A \delta + A \rho C \sum_{n=1}^\infty \frac{(-1)^n (\rho V)^{n-1} D_{n}}{n! D_{n+1}}. \ee
The infinite sum can be evaluated and is equal to
\be C \sum_{n=1}^\infty \frac{(-1)^n (\rho V)^{n-1} D_{n}}{n! D_{n+1}} \!= -\frac{1}{(d-1)!} \left(\frac{d}{2}H_\rho\!+\! 1\right)\!\left(\frac{d}{2}H_\rho\!+\!2\right)\cdots\left(\frac{d}{2}H_\rho\!+\!d\!-\!1\right)P, \ee
where $H_\rho := \rho \frac{\partial}{\partial \rho}$ is a differential operator.

For $d=2$ we have $A=2$ and
\be \Box P = 2\delta -2\rho(H_\rho+1)P.\ee
For $d=4$, we have $A=8\pi$ and
\be \label{eq:BoxP4d}\Box^2 P = 8\pi\delta -\frac{4\pi \rho}{3}(2H_\rho+1)(2H_\rho+2)(2H_\rho+3)P.\ee
This expression will be useful in section \ref{sec:dAlermbertianAnalysis}

\section{Propagators and d'Alembertians}
The function $P(x)$ plays a central role in a number of areas of causal set research. We now briefly describe two relevant areas.

\subsection{Propagators}

\label{sec:PropagatorIntro}

Propagators for the Klein-Gordon equation are solutions to the equation:
\be (\Box + m^2) G_m(x) = \delta(x) \ee
The \emph{retarded propagator} is the solution $G_m(x)$ that is supported only in the forward light cone $\Gamma_0^+$. In $d=4$ this is:
\be G_m(x) = C(x) \left( \frac{1}{2 \pi} \delta(x^2) - \frac{m}{4\pi} \frac{J_1\left(m \sqrt{x^2}\right)}{\sqrt{x^2}}\right),\ee
where $J_1$ is a Bessel function of the first kind.

In \cite{Johnston:2008} a causal set model for the retarded propagator was presented for $d=2$ and $d=4$. In $d=4$ this model involved a quantum mechanical path integral that summed amplitudes assigned to paths in the causal set. The expectation value of the propagator for a causal set generated by sprinkling into $\mink^4$ with density $\rho$ was equal to:
\be K_m(x): = \sum_{n=1}^\infty a^n b^{n-1} P_n(x)\quad \textrm{with}\quad a = \frac{\sqrt{\rho}}{2 \pi \sqrt{6}} \qquad b= -\frac{m^2}{\rho}. \ee
It was established in \cite{Johnston:2008} that
\be \label{eq:PropLimit}
\lim_{\rho\to\infty} K_m(x) = G_m(x). \ee
In particular, the massless propagator is related to $P$ through:
\be \lim_{\rho\to\infty} a P(x) =  G_0(x) = C(x) \frac{1}{2\pi} \delta(x^2) \ee

The results of this paper (section \ref{sec:PropagatorAnalysis}) provide the correction terms to \eqref{eq:PropLimit} for \emph{finite} sprinkling densities.

\subsection{Discrete d'Alembertians}

\label{sec:dAlembertianIntro}

In recent years a number of researchers have worked on causal set models that reproduce the behavior of the continuum d'Alembertian $\Box$ operator.

In these models, a field $\phi$ is represented by values assigned to each causal set element. By computing an alternating sum of these field values over layers of elements in the causal set the model is able to approximate the d'Alembertian operator applied to the field: $\Box \phi$.

More specifically, the expectation value of the models for sprinklings into $\mink^d$ with density $\rho$ is an integral kernel $\Bbar(x)$ that satisfies
\be \label{eq:dAlembertianLimit} \lim_{\rho \to \infty} \Bbar * \phi = \Box \phi\ee
for suitable functions $\phi(x)$. The kernels $\Bbar$ are the sum of a delta-function and the function $P$ acted upon with appropriate differential operators.

The first model, presented in \cite{Sorkin:2008}, gave, for $d=2$:
\be \bar{B} = 2\rho\delta - 2\rho^2(H_\rho+1)(H_\rho+2) P.\ee
where $H_\rho:= \rho \frac{\partial}{\partial \rho}$.

This was extended, in \cite[eq (4)]{Benincasa:2010}, to $d=4$:
\be \bar{B} = \frac{4\sqrt{\rho}}{\sqrt{6}}\left(\delta - \frac{\rho}{6} (2H_\rho+1)(2H_\rho+2)(2H_\rho+3) P \right).\ee
We note that equation \eqref{eq:BoxP4d} provides a surprisingly simple expression for $\bar{B}$ in $d=4$:
\be \label{eq:sweetysalty} \Box^2 P = \frac{2 \pi \sqrt{6}}{\sqrt{\rho}} \bar{B},\ee
This connection has been called the sweety-salty duality and will be used in section \ref{sec:dAlermbertianAnalysis}.

The models have been extended to higher dimensions \cite{Dowker:2013vba, Glaser:2014} as:
\be \bar{B} = \alpha_d \delta + \beta_d {\cal{O}}_d P, \ee
where, for even $d = 2n$ we have: \be \label{eq:Oop}{\cal{O}}_{2n} := \frac{1}{(n+1)!}\left(\frac{d}{2}H_\rho+1\right)\left(\frac{d}{2}H_\rho+2\right)\ldots\left(\frac{d}{2}H_\rho+n+1\right)\ee
and for odd $d=2n+1$ we have ${\cal{O}}_{2n+1} := {\cal{O}}_{2n}$. 
The $\alpha_d$ and $\beta_d$ are constants chosen to give the correct answers in $d$ dimensions. They are given in closed-form in \cite{Glaser:2014}.

Generalizations of these models have also been proposed recently \cite{Aslanbeigi:2014}. The d'Alembertian work has also been applied to model the scalar curvature of a causal set, as discussed in \cite{Benincasa:2010}.

The notation for this work sometimes use a sprinkling density $\rho$ as the variable, sometimes a discreteness length scale $l:= \rho^{-d}$. We have therefore clarified our $H$ operators with a subscript. In \cite{Dowker:2013vba, Glaser:2014} they use a differential operator $H_l := -l \frac{\partial}{\partial l}$ which is related to ours as $H_l = d H_\rho$. The sign of $\Bbar$ also depends on the metric signature used for $\mink^d$.
In addition, the work is often phrased using $P(-x)$, a function supported on past light cone, rather than the future light cone. We hope the conventions used here are clear.

The results of section \ref{sec:dAlermbertianAnalysis} give the correction terms to the limit in \eqref{eq:dAlembertianLimit}. Previous approaches to obtaining the correction terms involve Taylor-expanding the fields $\phi(x)$ and calculating the integral $\Bbar * \phi$ term-by-term. Each of these integrals typically diverge so are evaluated with a large-scale cut-off. This cut-off breaks the Lorentz-invariance of the theory so is then removed with care. The methodology used in \ref{sec:dAlermbertianAnalysis} is manifestly Lorentz-invariant and gives the correction terms without direct reference to the field $\phi$.

\section{\texorpdfstring{Large $\rho$ expansion}{Large rho expansion}}
\label{sec:LargeRho}
The models just described both require taking the infinite $\rho$ limit to give agreement with the continuum quantities. The central premise of causal set theory is that spacetime \emph{is} fundamentally discrete. As such, the most interesting regime is when $\rho$ is large but \emph{finite}. Indeed, we would expect $\rho$ to be sufficiently large so that discreteness effects have gone undetected so far. Often the discreteness scale is assumed to be based on the Planck-length. To sprinkle into $\mink^4$ with a Planckian density we could take $\rho$ to be the inverse of the Planck 4-volume: $\rho =c^7/(G \hbar)^2$ which corresponds to $4.4 \times 10^{147}$ causal set elements per cubic-metre-second of spacetime --- certainly a large number.

We now look at the large $\rho$ behaviour of $P(x)$. We will then apply these results to look at the large $\rho$ behaviour of the propagator and d'Alembertian models. To get started, we summarize two tools we shall use extensively: the Riesz distributions and the Laplace transform.

\subsection{Riesz Distributions}

An important class of Lorentz invariant retarded distributions are the Riesz distributions (see \cite{Kolk:Riesz} for full review). These are related to powers of the quadratic form $\omega(x) = x^2$ and have an action on test functions $f$ defined by:
\be \langle R_\alpha, f\rangle := \frac{1}{H_d(\alpha)} \int_{\Gamma^+_0} \omega^{\frac{\alpha-d}{2}}(x) f(x) \mathrm{d}^{d} x \ee
where $H_d(\alpha)$ is a normalization constant:
\be H_d(\alpha) = \pi^{\frac{d-2}{2}} 2^{\alpha-1} \Gamma\left(\frac{\alpha}{2}\right)\Gamma\left(\frac{\alpha -d + 2}{2}\right).\ee
With this normalization, the map $\alpha \mapsto R_\alpha$ is a complex-analytic mapping with the following properties:
\be R_\alpha * R_\beta = R_{\alpha + \beta}, \ee
\be R_0 = \delta, \ee
\be \Box^k R_{\alpha+2k} = R_{\alpha}. \ee
We note that the expected number of chains, defined in \eqref{eq:ChainForm} and treated as a distribution, is related to the $R_\alpha$ through $C_n = \rho^{n-1} (H_d(d))^n R_{n d}$.

Another important distribution is the infinitesimal generator of $R_\alpha$ given by:
\be \tau := {{\frac{d}{d\alpha}}\vline}_{\alpha=0} R_{\alpha} \ee
This has been analyzed thoroughly in the classification of Lorentz invariant distributions supported on the forward light cone \cite{Kolk:Lorentz, Kolk:Riesz}. For our purposes, it will also be useful to define a related distribution $\tau'[A]$, using an $\alpha$-independent constant $A$, as:
\be  \label{eq:TauDef} \tau'[A] := {{\frac{d}{d\alpha}}\vline}_{\alpha=0} A^{\alpha/2} R_{\alpha} = \frac{1}{2} \ln(A) \delta + \tau\ee
	The distribution $\tau$ is defined up to the addition of multiples of the delta function \cite[Sec 7.2]{Kolk:Lorentz} so explicit spacetime expressions can only be given for test functions that vanish at 0. It is related to the pullback of derivatives of the delta function on $\mathbb{R}$ through \cite[eq 6.3]{Kolk:Riesz}:
\be \tau = \pi^{-\frac{d-2}{2}} \omega^*(\delta_0^{\left(\frac{d-2}{2}\right)}). \ee
In $d=4$, we have the following spacetime expression for $\tau$ \cite[sec 1.0, Remark 2]{Kolk:Riesz}, \cite[sec 7.2]{Kolk:Lorentz}:
 \be \langle \tau, f \rangle = \frac{1}{4\pi} \int_{\mathbb{R}^3} f(|x|,x) \frac{\mathrm{d}^3x}{|x|^3} - \frac{1}{4 \pi} \int_{\mathbb{R}^3} \frac{\partial f}{\partial x_0}(|x|,x) \frac{\mathrm{d}^3x}{|x|^2} \ee
For even dimensions $d$, we have (compare \cite[p283]{Kolk:Riesz}):
\be \label{eq:LogTau} \langle \tau, f \rangle = {{\frac{d}{d\alpha}}\vline}_{\alpha=0} \langle  R_{\alpha}, f \rangle = {{\frac{d}{d\alpha}}\vline}_{\alpha=0} \langle  R_{\alpha+d},\Box^{d/2} f \rangle =  \frac{1}{2 H_d(d)} \int_{\Gamma^+_0} \Box^{d/2} f(x) \ln( w(x)) \mathrm{d}^dx\ee
From the properties of $R_\alpha$, we have $R_{-d} * R_d = R_0$ which, in other notation, is \be \frac{1}{H_d(d)} C * \Box^{d/2} \delta = \delta\ee
which gives:
\be \label{eq:Deltalightcone} \delta * f = \frac{1}{H_d(d)} C * \Box^{d/2} \delta *f = \frac{1}{H_d(d)} C * \Box^{d/2} f = \frac{1}{H_d(d)} \int_{\Gamma^+_0} \Box^{d/2} f(x) \mathrm{d}^dx\ee
Combining this with \eqref{eq:TauDef} and \eqref{eq:LogTau} gives:
\be \langle \tau'[A], f \rangle = \frac{1}{2H_d(d)} \int_{\Gamma^+_0} \Box^{d/2} f(x) \ln(A w(x)) \mathrm{d}^dx \ee

\subsection{Laplace Transform}
\label{sec:Laplace}

Following \cite[I,1;1]{Laplace}, the Laplace transform of a function $f(x)$ defined on $\mink^d$ is:
\be L\{f\} := \int_{\mink^d} e^{-i\langle x, z \rangle} f(x) \mathrm{d}^dx, \ee
where $z=(z_0,z_1,\ldots,z_{d-1}) \in \mathbb{C}^d$ and $\langle x, z\rangle = x_0 z_0 + x_1 z_1 + \ldots + x_{d-1} z_{d-1}$. The Fourier transform of $f(x)$ can be obtained from the Laplace transform in the limit as the imaginary part of $z$ tends to zero.

The Laplace transform converts convolutions into products:
\be L\{f*g\} = L\{ f\}  L\{ g\},  \ee
and satisfies
\be L\{\Box f\} = s L\{f\}, \ee
where $s = - z^2 = z_1^2 + z_2^2 + \ldots + z_{d-1}^2 - z_0^2$.

The Laplace transform of the Riesz distributions is \cite[II,3;3]{Laplace}:
\be L\{ R_\alpha \} = s^{-\alpha/2}, \ee
which implies
\be L\{ \tau \} = {{\frac{d}{d\alpha}}\vline}_{\alpha=0} L\{ R_\alpha\} = -\frac{1}{2}\ln(s),\ee
and
\be \label{eq:LaplaceTau} L\{ \tau'[A] \} = {{\frac{d}{d\alpha}}\vline}_{\alpha=0} L\{ A^{\alpha/2} R_\alpha\} = {{\frac{d}{d\alpha}}\vline}_{\alpha=0} \left(\frac{A}{s}\right)^{\alpha/2} = \frac{1}{2}\ln\left(\frac{A}{s}\right).\ee

The Laplace transform of retarded Lorentz invariant functions supported on the forward light cone has been obtained in \cite{Laplace}. We can apply this expression to $P$ by recognizing that, for $x \in \Gamma^+_0$, $P(x)$ is equal to:
\be P(x) = \exp\left(-c_d \rho \omega(x)^{d/2}\right) \ee
 (compare \eqref{eq:Vdef} and \eqref{eq:Pdef}). From \cite[I,2;1]{Laplace} the Laplace transform of $P$ is therefore given by the one-dimensional integral:
\be \label{eq:ltrans} L\{P\} = \frac{(2\pi)^{(d-2)/2}}{s^{(d-2)/4}} \int_0^\infty \exp\left( -c_d \rho \lambda^{d/2}\right) \lambda^{(d-2)/4} K_{(d-2)/2}(\sqrt{\lambda s}) \mathrm{d}\lambda \ee
where $K_n(t)$ is a modified Bessel function of the second kind.

This will be the central equation which will enable us to find the large $\rho$ behaviour of $P$. We split the discussion depending on whether $d$ is even or odd. 

We note that \eqref{eq:ltrans} also appeared in \cite[eq 3.7]{Aslanbeigi:2014} where it was used to examine at the behaviour of $\LP$ for large $s$.

\subsection{Even Dimensions}

For $d$ even, it is useful to evaluate \eqref{eq:ltrans} with the substitution $t = \lambda s/4$ to get:
\be \label{eq:ltransxeven} L\{P\} = \frac{4 (4\pi)^{(d-2)/2}}{s^{d/2}} 
\int_0^\infty \exp\left(-c_d \rho {\left(\frac{4}{s}\right)}^{d/2} t^{d/2}\right) t^{(d-2)/4} K_{(d-2)/2}(2\sqrt{t}) \mathrm{d}t \ee

For $d=2n+2$ ($n=0,1,\ldots$), the Bessel function in \eqref{eq:ltransxeven} has integer order $n$ with a series expansion of the form \cite[8.446]{integrals}:
\begin{align}  \label{eq:BesselExpand}t^{n/2} K_n(2\sqrt{t}) &= \frac{1}{2} \sum_{k=0}^{n-1} (-1)^k \frac{(n-k-1)!}{k!}t^k \\&+ (-1)^{n+1}\sum_{k=0}^\infty \frac{t^{n+k}}{2 (k!) (n+k)!}\left[ \ln(t) - \psi(k+1) - \psi(n+k+1)\right] \nonumber\end{align}
where $\psi(k)$ is the digamma function \cite[8.330]{integrals}. This satisfies $\psi(1) = -\gamma$ and $\psi(k+1) = \psi(k) + 1/k$ where $\gamma$ is the Euler-Mascheroni constant.

Substituting the Bessel function expansion \eqref{eq:BesselExpand} into \eqref{eq:ltransxeven} results in an infinite series with terms of the form:
\be A(a,d,k) := \int_0^\infty \exp(-a t^{d/2}) t^{k}\mathrm{d}t,\ee
and
\be B(a,d,k) := \int_0^\infty \exp(-a t^{d/2}) t^{k} \ln(t) \mathrm{d}t.\ee
These integrals can be evaluated as:
\be A(a,d,k) = \frac{2}{d} \, a^{-\frac{2 \, {\left(k + 1\right)}}{d}} \Gamma\left(\frac{2 \, {\left(k + 1\right)}}{d}\right),\ee
\be B(a,d,k) = -\frac{4}{d^2} \, a^{-\frac{2 \, {\left(k + 1\right)}}{d}} \Gamma\left(\frac{2 \, {\left(k + 1\right)}}{d}\right) {\left(\ln\left(a\right) - \psi\left(\frac{2 \, {\left(k + 1\right)}}{d}\right)\right)}.  \ee
Combining these gives an infinite series for $L\{P\}$ of:
\begin{align}
   \LP &= \frac{4(4\pi)^n}{s^{n+1}} \Bigg(\frac{1}{2} \sum_{k=0}^{n-1} (-1)^k \frac{(n-k-1)!}{k!}A\left(a,d,k\right) 
	\\  &+ (-1)^{n+1}\sum_{k=0}^\infty \frac{1}{2 (k!) (n+k)!} [  B\left(a,d,n+k\right) - (\psi(k+1) + \psi(n+k+1)) A\left(a,d,n+k\right) ]  \Bigg) \nonumber
	\end{align}
where $n=(d-2)/2$ and $a = c_d \rho \left(4/s\right)^{d/2}$.

\subsubsection{Dimension $d=2$}
For $d=2$, we have $n=0$ and:
\be \label{eq:Bessel0} K_0(2\sqrt{t}) = \sum_{k=0}^\infty \frac{t^{k}}{2(k!)^2}\left[ -\ln(t) + 2\psi(k+1)\right] \ee
\be a = \frac{2 \rho}{s} \ee
\be A(a,2,k) = \frac{\Gamma(k+1)}{a^{k+1}} \ee
\be B(a,2,k) = -\frac{\Gamma(k+1)}{a^{k+1}}\left(\ln\left(a\right) - \psi\left(k + 1\right) \right) \ee
Combining these gives:
\begin{align} \label{eq:LP2d}  L\{P\} &= \sum_{k=0}^\infty \frac{4}{2s(k!)^2}\left[ -B(a,2,k) + 2\psi(k+1)A(a,2,k)\right] \\
 &= 2 \sum_{k=0}^\infty \frac{s^{k} }{ k!} \frac{1}{(2\rho)^{k+1}}\left[\ln\left(\frac{2 \rho}{s}\right) + \psi\left(k + 1\right) \right] \end{align}
The first few terms are, in descending powers of $\rho$ (noting that $\ln(\rho)/\rho^{k+1} = O(1/\rho^k)$):
\be L\{P\} = \frac{1}{2\rho}\ln\left(\frac{2 \rho}{s}\right) + \frac{\psi(1)}{\rho} +\frac{s}{2\rho^2} \ln\left(\frac{2 \rho}{s}\right) + \frac{s \psi(2)}{2\rho^2} + \frac{s^2}{8\rho^3} \ln\left(\frac{2\rho}{s}\right) + O\left(\frac{1}{\rho^3}\right)\ee
Translating this into position space, using the results of section \ref{sec:Laplace}, gives
\be P = \frac{1}{\rho}\tau'[2\rho] + \frac{\psi(1)}{\rho} \delta + \frac{1}{\rho^2} \Box\tau'[2\rho] + \frac{\psi(2)}{2\rho^2} \Box \delta + \frac{1}{4\rho^3} \Box^2 \tau'[2\rho] + O\left(\frac{1}{\rho^3}\right)\ee
We also note that $\LP$ can be evaluated in closed-form as:
\be L\{P\} = \frac{1}{\rho} \exp\left(\frac{s}{2\rho}\right) E_1\left(\frac{s}{2\rho}\right) \ee
where $E_1$ is the Exponential Integral.

\subsubsection{Dimension $d=4$}
For $d=4$, we have $n=1$ and:
\be \label{eq:Bessel1} \sqrt{t} K_1(2\sqrt{t}) = \frac{1}{2} + \sum_{k=0}^\infty \frac{t^{1+k}}{2 (k)!(1+k)!}\left[ \ln(t) - \psi(k+1) - \psi(k+2)\right] \ee
\be a = \frac{2 \pi \rho}{3 s^2} \ee
\be A(a,4,0) = \frac{\sqrt{\pi}}{2\sqrt{a}}\ee
\be A(a,4,k) = \frac{1}{2}a^{-\frac{k + 1}{2}} \Gamma\left(\frac{{\left(k + 1\right)}}{2}\right)\ee
\be B(a,4,k) = -\frac{1}{4} a^{-\frac{k + 1 }{2}} {\left(\ln\left(a\right) - \psi\left(\frac{\left(k + 1\right)}{2}\right)\right)} \Gamma\left(\frac{ k + 1 }{2}\right) \ee
Combining these gives:
\begin{multline}\label{eq:LP4d}
   L\{P\} = \frac{16\pi}{s^2} \Bigg(
	\frac{A(a,4,0)}{2}  
	 + \sum_{k=0}^\infty \frac{1}{2 (k!) (1+k)!}
	\left[  B(a,4,1+k) - (\psi(k+1) + \psi(k+2))A(a,4,1+k)\right] \Bigg) \\
= \frac{16\pi}{s^2} \left(
	 \frac{\sqrt{\pi}}{4\sqrt{a}}  + \sum_{k=0}^\infty \frac{a^{-\frac{k+2}{2}} \Gamma\left(\frac{k+2}{2}\right)}{8 (k!) (k+1)!}\left[ \psi\left(\frac{k+2}{2}\right) - \ln(a) - 2\psi(k+1) - 2\psi(k+2)\right] \right) \\
  = \frac{2 \pi \sqrt{6}}{s \sqrt{\rho}} + 6\pi \sum_{k=0}^\infty \frac{(\sqrt{3} s)^{k} \Gamma\left(\frac{k+2}{2}\right)}{(k!) (k+1)! \left(\sqrt{2 \pi \rho}\right)^{k+2}}
    \left[ \psi\left(\frac{k+2}{2}\right) - \ln\left( \frac{2 \pi \rho}{3 s^2}\right) - 2\psi(k+1) - 2\psi(k+2)\right] \end{multline}
The first few terms, in descending powers of $1/\sqrt{\rho}$, are:
\be \LP =\frac{2 \pi \sqrt{6}}{s \sqrt{\rho}} - \frac{3}{ \rho}\ln\left( \frac{2 \pi \rho}{3 s^2}\right) - \frac{3(2 - 3\gamma)}{ \rho} - \frac{3\sqrt{3}s}{4\sqrt{2}\sqrt{\rho}^3} \ln\left( \frac{2 \pi \rho}{3 s^2}\right) + O\left(\frac{1}{\sqrt{\rho}^3}\right)\ee
where $\gamma$ is the Euler-Mascheroni constant.

Translating this into position space, using the results of section \ref{sec:Laplace}, gives
\be P = \frac{2 \pi \sqrt{6}}{\sqrt{\rho}} R_{2} - \frac{12}{ \rho} \tau'\left[\left(\frac{2 \pi \rho}{3}\right)^{1/2}\right] - \frac{3(2 - 3\gamma)}{ \rho}\delta - \frac{3\sqrt{3}}{\sqrt{2}\sqrt{\rho}^3} \Box \tau'\left[\left(\frac{2 \pi \rho}{3}\right)^{1/2}\right] + O\left(\frac{1}{\sqrt{\rho}^3}\right),\ee
where $R_{2} = G_0$ is the massless retarded propagator: $\Box R_{2} = \delta$.

\subsection{Odd Dimensions}

For $d$ odd, it is useful to evaluate \eqref{eq:ltrans} with the substitution $t = \sqrt{\lambda s}$ to get:
\be \label{eq:ltransxodd} L\{P\} = \frac{2 (2\pi)^{(d-2)/2}}{s^{d/2}} 
\int_0^\infty \exp\left(-c_d \rho {\left(\frac{1}{s}\right)}^{d/2} t^{d}\right) t^{d/2} K_{(d-2)/2}(t) \mathrm{d}t. \ee

For $d=2n+3$ ($n=0,1,\ldots$), the Bessel function in \eqref{eq:ltransxodd} is of half-integer order $n+\frac{1}{2}$. These can be expressed in terms of exponentials and powers  to give \cite[8.468]{integrals}:
\be t^{d/2} K_{n+\frac{1}{2}}(t) = \sqrt{\frac{\pi}{2}} \exp(-t) \sum_{k=0}^n \frac{(n+k)!}{k! (n-k)! 2^k} t^{n-k+1}. \ee
Combining this with \eqref{eq:ltransxodd} gives $L\{P\}$ as:
\be \LP = \frac{(2\pi)^{(d-1)/2}}{s^{d/2}} \sum_{k=0}^n \frac{(n+k)!}{k! (n-k)! 2^k}
\int_0^\infty \exp\left(-c_d \rho {\left(\frac{1}{s}\right)}^{d/2} t^{d}\right)  \exp(-t) t^{n-k+1}  \mathrm{d}t, \ee
with $n=(d-3)/2$. By expanding the $\exp(-t)$ in the integrand, an infinite series for $\LP$ can be generated in powers of ${\rho}^{-2/d}$:
\begin{align} \LP &= \frac{(2\pi)^{(d-1)/2}}{s^{d/2}} \sum_{m=0}^\infty \sum_{k=0}^n \frac{(-1)^m (n+k)!}{m! k! (n-k)! 2^k}
\int_0^\infty \exp\left(-c_d \rho {\left(\frac{1}{s}\right)}^{d/2} t^{d}\right)  t^{m+n-k+1}  \mathrm{d}t \\
&=\frac{(2\pi)^{(d-1)/2}}{s^{d/2}} \sum_{m=0}^\infty \sum_{k=0}^n \frac{(-1)^m(n+k)!}{m!k! (n-k)! 2^k}
A(a,2d,m+n-k+1),
\end{align}
where $a = c_d\rho (1/s)^{d/2}$.

\subsubsection{Dimension $d=3$}

For $d=3$, we have $n+\frac{1}{2}=\frac{1}{2}$ and:
\be t^\frac{3}{2} K_\frac{1}{2}(t) = \sqrt{\frac{\pi}{2}} t \exp(-t) = \sqrt{\frac{\pi}{2}} \sum_{k=0}^{\infty} \frac{(-1)^k t^{k+1}}{k!}\ee
\be a = \frac{\pi\rho}{12s^{3/2}} \ee
which combine to give
\begin{align} \LP &= \frac{2\pi}{s^{3/2}} 
\int_0^\infty \exp\left(-a t^3\right)  \exp(-t) t\,  \mathrm{d}t \\
&= \frac{2\pi}{s^{3/2}} \sum_{m=0}^\infty \int_0^\infty \exp\left(-a t^3\right) \frac{(-1)^m t^{m+1}}{m!}  \mathrm{d}t
=\frac{2\pi}{s^{3/2}} \sum_{m=0}^\infty \frac{(-1)^m A(a, 6, m+1)}{m!} \nonumber
\end{align}
we have
\be A(a,6,m+1) = \frac{1}{3} \, a^{-\frac{ {\left(m + 2\right)}}{3}} \Gamma\left(\frac{ { m + 2 }}{3}\right)\ee
therefore 
\be \LP = \frac{2\pi}{3s^{3/2}} \sum_{m=0}^\infty  \left(\frac{12 s^{3/2}}{\pi \rho}\right)^{\frac{m + 2}{3}} \frac{(-1)^m}{m!}\Gamma\left(\frac{m + 2}{3}\right)\ee
The first few terms, in descending powers of $\rho^{1/3}$ are:
\be L\{P\} = \frac{2\pi}{3s^{1/2}} \left(\frac{12 }{\pi \rho}\right)^{\frac{2}{3}} \Gamma\left(\frac{2}{3}\right) + \frac{8}{ \rho} +  \frac{2\pi s^{1/2}}{3}\left(\frac{12}{\pi \rho}\right)^{\frac{4}{3}} \Gamma\left(\frac{4}{3}\right) + \frac{2\pi s}{3}\left(\frac{12}{\pi \rho}\right)^{\frac{5}{3}} \Gamma\left(\frac{5}{3}\right) 
+ O\left(\frac{1}{\rho^2} \right)
\ee
Translating to position space we have:
\be L\{P\} = \frac{2\pi}{3}  \left(\frac{12 }{\pi \rho}\right)^{\frac{2}{3}} \Gamma\left(\frac{2}{3}\right)R_1 + \frac{8}{ \rho}\delta +  \frac{2\pi }{3} \left(\frac{12}{\pi \rho}\right)^{\frac{4}{3}} \Gamma\left(\frac{4}{3}\right)R_{-1} + \frac{2\pi }{3}  \left(\frac{12}{\pi \rho}\right)^{\frac{5}{3}} \Gamma\left(\frac{5}{3}\right) R_{-2}
+ O\left(\frac{1}{\rho^2} \right)
\ee

\section{Analysis}

\label{sec:Analysis}

Using the large $\rho$ expansions for $P$ we can find the large $\rho$ behaviour of the propagators and discrete d'Alembertians discussed in sections \ref{sec:PropagatorIntro} and \ref{sec:dAlembertianIntro}.

\subsection{Propagators}

\label{sec:PropagatorAnalysis}
Recall that the expectation value of the causal set retarded propagator is:
\be K_m: = \sum_{n=1}^\infty a^n b^{n-1} P_n,\qquad a = \frac{\sqrt{\rho}}{2 \pi \sqrt{6}} \qquad b= -\frac{m^2}{\rho} \ee
with the massless propagator $K_0 := a P$.

We note that the continuum retarded propagator $G_m$ has Laplace transform:
\be L\{G_m\} = \frac{1}{s + m^2}\ee
and
\be L\{\Box G_m\} = \frac{s}{s + m^2}\ee

The series \eqref{eq:LP4d} shows that:
\be \lim_{\rho \to \infty} L\{K_0 \}  = \lim_{\rho \to \infty} L\left\{ \frac{\sqrt{\rho}}{2 \pi \sqrt{6}} P \right\} = \frac{1}{s} = L\{G_0\}\ee
as was shown, via Fourier transform, in \cite[eq 3.39]{Johnston:2008}.

We can summarize \eqref{eq:LP4d} as
\be \label{eq:K0series} L\{K_0\} = \sum_{k=0}^{\infty} \frac{A_k}{(\sqrt{\rho})^k} = A_0 + \frac{A_1}{\sqrt{\rho}} + \frac{A_2}{\sqrt{\rho}^2} + \ldots \ee
where
\be A_0 = \frac{1}{s} \ee
\be A_{k+1} =  \frac{3(\sqrt{3} s)^{k} \Gamma\left(\frac{k+2}{2}\right)}{\sqrt{6}(k!) (k+1)! \left(\sqrt{2 \pi}\right)^{k+2}}
    \left[ \psi\left(\frac{k+2}{2}\right) - \ln\left( \frac{2 \pi \rho}{3 s^2}\right) - 2\psi(k+1) - 2\psi(k+2)\right]\ee
We then have:
\be \label{eq:Kmseries} L\{K_m\} = \frac{a\LP}{1-\rho ab\LP} = \frac{L\{K_0\}}{1+m^2 L\{K_0\}} \ee
We can treat \eqref{eq:K0series} as a formal power series in $1/\sqrt{\rho}$ and then evaluate \eqref{eq:Kmseries} through manipulation of power series.

In general we have, by applying \cite[0.313]{integrals}:
\be \label{eq:Reciprocal} F(x) = \sum_{k=0}^\infty a_k x^k \implies \frac{F(x)}{1-\lambda F(x)} = \frac{1}{1-\lambda a_0} \sum_{k=0}^\infty b_k x^k\ee
with
\be b_0 = a_0 \quad b_n = a_n + \frac{\lambda}{1-\lambda a_0}\sum_{k=1}^n   a_k b_{n-k}. \ee

Therefore, treating $L\{K_0\}$ as a formal power series in $1/\sqrt{\rho}$ we can use \eqref{eq:Reciprocal} with $\lambda = -m^2$ to get:
\be L\{K_m\} = \frac{L\{K_0\}}{1+m^2 L\{K_0\} } = \frac{1}{1+m^2 A_0} \sum_{k=0}^\infty \frac{B_k}{(\sqrt{\rho})^k} \ee
with
\be B_0 = A_0 \quad B_n = A_n - \frac{m^2}{1 + m^2 A_0}\sum_{k=1}^n A_k B_{n-k}. \ee
We have, for example:
\be B_1 = A_1 - \frac{m^2 A_1 A_0}{1 + m^2 A_0} =  \frac{A_1}{1 + m^2 A_0}\ee
If we recognize that:
\be \frac{A_0}{1 + m^2 A_0}= \frac{\frac{1}{s}}{1 + \frac{m^2}{s}} = \frac{1}{s + m^2} = L\{G_m\}\ee
\be \frac{1}{1 + m^2 A_0}= \frac{1}{1 + \frac{m^2}{s}} = \frac{s}{s + m^2} = L\{\Box G_m\}\ee
\be A_{1} =  \frac{3}{2 \pi \sqrt{6} }
    \left[ 3\gamma - 2 - \ln\left( \frac{2 \pi \rho}{3 s^2}\right) \right]\ee

we see that:
\be L\{K_m\} =  \frac{1}{s + m^2} + 
\frac{s^2}{(s+m^2)^2} \frac{3}{2 \pi \sqrt{6} }
    \left[ 3\gamma - 2 - \ln\left( \frac{2 \pi \rho}{3 s^2}\right) \right] \frac{1}{\sqrt{\rho}} + \ldots
\ee
Translating this into position space and keeping the leading-order terms, this is:

\be K_m = G_m + \frac{3}{2 \pi \sqrt{6} } \Box G_m * \Box G_m * \left[(3\gamma - 2)\delta -  2\tau'\left[\left(\frac{2 \pi \rho}{3}\right)^{1/2} \right]\right] \frac{1}{\sqrt{\rho}} + \ldots \ee

\subsection{d'Alembertians}

\label{sec:dAlermbertianAnalysis}

We can phrase the discussion in section \ref{sec:dAlembertianIntro} in terms of distributions and the Laplace transform. The goal of the work is to show that the integral kernel $\Bbar$ satisfies $\lim_{\rho\to\infty} \bar{B} = \Box \delta$. Taking the Laplace transform this becomes:
\be \label{eq:bbarlaplace} \lim_{\rho \to \infty} L\{ \bar{B} \} = s.\ee
The causal set d'Alembertian operator in $\mink^d$ is of the form $\Bbar = \alpha_d \delta + \beta_d{\cal{O}}_d P$ for appropriate $\alpha_d$, $\beta_d$ (given in \cite{Glaser:2014}) and ${\cal{O}}_d$ given in \eqref{eq:Oop}. 

We can use the large $\rho$ expansion for $\LP$ to obtain the general correction terms for $L\{\Bbar\}$.
We note that the terms in the expansion of $\LP$ are of the form \be \frac{ \ln(A \rho)+B}{\rho^k}\ee for $\rho$-independent values $A$ and $B$. A simple calculation gives
\be (H_\rho + n) \left(\frac{\ln(A \rho) + B}{\rho^k} \right)=  \frac{(n-k)(\ln(A \rho)+B)+1}{\rho^k}. \ee
We note that the method outlined here to calculate the d'Alembertian corrections is Lorentz-invariant and does not depend on a large-scale cut-off, or on properties of the field for which the d'Alembertian is being calculated. This is an improvement on the correction terms calculated in previous work \cite{Dowker:2013vba}.

We shall describe the details for the $d=2$ and $d=4$ cases.

\subsubsection{Dimension $d=2$}
In $d=2$ we have
\be \LP = 2 \sum_{k=0}^\infty \frac{s^{k} }{ k!} \frac{1}{(2\rho)^{k+1}}\left[\ln\left(\frac{2 \rho}{s}\right) + \psi\left(k + 1\right) \right]
\ee
and
\be \bar{B} = 2 \rho \delta - 2\rho^2 (H_\rho+2) (H_\rho+1)P.\ee
Combining these gives
\be L\{\bar{B}\} = 
s -  4\rho^2 \sum_{k=2}^\infty \frac{s^{k} }{ k!} \frac{1}{(2\rho)^{k+1}}\left[-k(1-k)\left(\ln\left(\frac{2 \rho}{s}\right) + \psi\left(k + 1\right)\right) + 1 - 2k \right].\ee
The first few terms of this are
\be  L\{\bar{B}\} = s - \frac{s^2}{2\rho} \left(\ln\left(\frac{2 \rho}{s}\right) - \gamma \right) +  \ldots\ee
which validates that \eqref{eq:bbarlaplace} is satisfied and gives the first-order corrections.

\subsubsection{Dimension $d=4$}

The most physically-interesting dimension is $d=4$. From \eqref{eq:BoxP} and \eqref{eq:sweetysalty} the series for $\bar{B}$ can be given immediately by as:
\be \bar{B} = \frac{\sqrt{\rho}}{2 \pi \sqrt{6}}\Box^2 P \ee
or, taking Laplace transforms,
\be L\{\bar{B}\} = \frac{\sqrt{\rho}}{2 \pi \sqrt{6}} s^2 L\{P\} = s^2 L\{K_0\} = s^2 \sum_{k=0}^\infty \frac{A_k}{\sqrt{\rho}^k}\ee
where we have used \eqref{eq:K0series}. The first few terms are
\be L\{\bar{B}\} = s  - \frac{3}{2 \pi \sqrt{6} }
    \left[ 3\gamma - 2 - \ln\left( \frac{2 \pi \rho}{3 s^2}\right) \right] \frac{1}{\sqrt{\rho}}+ \ldots \ee
which validates that \eqref{eq:bbarlaplace} is satisfied and gives the first-order corrections.

\section{Conclusions and Future Work}

We have used the Laplace transform to obtain the large $\rho$ behaviour of the expectation values of the causal set retarded propagator and d'Alembertian models.

The work presented here focused on the expectation value of the causal set quantities. It is also important to understand the behaviour of the variance of these quantities and how this behaves in the large $\rho$ limit. This is a difficult question to address analytically. 
Numerical simulations \cite[sec 3.3]{Johnston:2008} suggest that the variance of the retarded propagators models tends to zero for large $\rho$. This may be expected since the causal set model sums amplitudes over the large number of paths in the causal set, leading to cancellations of random fluctuations.
Numerical simulations of the d'Alembertian operators suggest that their variance grows with the sprinkling density \cite{Sorkin:2008}. These fluctuations have been dampened by the introduction of a non-locality scale larger than the discreteness scale. The techniques presented here may be applicable to the large $\rho$ correction terms for these non-local d'Alembertian models.

The techniques presented here may be helpful in obtaining correction terms for related causal set models. The causal set Feynman propagator presented in \cite{Johnston:2009} is derived directly from the causal set retarded propagator through a matrix eigendecomposition. If the large $\rho$ behaviour of the Feynman propagator can be determined it would be a step closer to a realistic model for matter on a causal set spacetime background. The correction terms to the continuum values would be directly related to the scale of spacetime discreteness and could potentially offer experimental tests that compare the continuum and discrete theories. The results given here have included both the position space and Laplace-transform behaviour of the models. Quantum field theory calculations are often done in momentum space, using the Fourier transform. It is possible that the Laplace transform correction terms could be used to calculate directly in momentum space, with the $\tau'[A]$ distributions leading to terms logarithmic in the momentum and sprinkling density.
The scalar curvature model presented in \cite{Benincasa:2010} is derived directly from the $d=4$ d'Alembertian model. As such, it may be possible to determine its large $\rho$ behaviour from the results presented here.

In \cite{Dowker:2013vba, Glaser:2014} expressions for the d'Alembertian expectation values are given for arbitrary dimensions. Further work would also be to validate these results and obtain the general expression for the correction terms.

The methods described here may also help to obtain a clearer understanding of the expected number of paths for sprinklings into $\mink^d$ (specifically the behaviour of the $P_n(y-x)$ functions). How these depend on $V(y-x)$, the spacetime volume of a causal interval $[x,y]$, could provide kinematical information that characterizes causal sets that well-approximate Lorentzian manifolds. These could offer further ways to detect manifold-like causal sets based on statistics of paths (not chains) to find manifold quantities such as proper-time and spacetime dimension.

Although the Laplace transform requires the symmetries of Minkowski spacetime in its definition, it is possible that the position-space correction terms still apply for causal sets generated by sprinklings into curved Lorentzian manifolds. In particular, if the Riesz distributions $R_\alpha$ can be defined in curved Lorentzian manifolds then the $\tau$ and $\tau'[A]$ distributions can be defined by taking the  derivative of $R_\alpha$ with respect to $\alpha$.

\end{document}